\documentclass[apj]{emulateapj}
\newcommand{\Lsun}{L_{\odot}}        
\newcommand{\Msun}{M_{\odot}}
\newcommand{\ergs}{\rm erg~s^{-1}}
\newcommand{\OIII}{$[\rm O~ \sc{III}]$}

\newcommand{\kms}{\ifmmode {\rm km\ s}^{-1} \else km s$^{-1}$\ \fi}

\begin{document}

\title{Active Galactic Nuclei with Double-Peaked Balmer
Lines: I. Black Hole Masses and the Eddington ratios}

\author{Wei-Hao Bian \altaffilmark{1,}\altaffilmark{2}, Yan-Mei Chen
\altaffilmark{1}, Qiu-Sheng Gu\altaffilmark{3}, and Jian-Min Wang
\altaffilmark{1}} \altaffiltext{1}{Key Laboratory for Particle
Astrophysics, Institute of High Energy Physics, Chinese Academy of
Sciences, Beijing 100039, China} \altaffiltext{2}{Department of
Physics and Institute of Theoretical Physics, Nanjing Normal
University, Nanjing 210097, China; whbian@njnu.edu.cn}
\altaffiltext{3}{Department of Astronomy, Nanjing University,
Nanjing 210093, China}

\shorttitle{SMBH Masses and Eddington Ratios of Double-Peaked
AGNs}

\shortauthors{Bian, Chen, Gu \& Wang}

\slugcomment{Submitted to ApJ}

\begin{abstract}
Using the stellar population synthesis, we model the stellar
contribution for a sample of 110 double-peaked broad-lines AGNs
from the Sloan Digital Sky Survey (SDSS). The stellar velocity
dispersions ($\sigma_*$) are obtained for 52 double-peaked AGNs
with obvious stellar absorption features, ranging from 106 to 284
\kms. We also use multi-component profiles to fit \OIII
$\lambda\lambda4959,5007$ and H$\beta$ emission lines. Using the
well-established $M_{\rm bh}-\sigma_*$ relation, the black hole
masses are calculated to range from $1.0\times 10^{7}$ to
$5.5\times 10^{8}$ $\Msun$, and the Eddington ratio from about
0.01 to about 1. Comparing with the known $R_{\rm BLR}-L$
relation, we can get the factor $f$, which indicates BLRs'
geometry, inclination and kinematics. We find that $f$ far
deviates from 0.75, suggesting the non-virial dynamics of broad
line regions. The peak separation is mildly correlated with the
Eddington ratio and SMBH mass with almost the same correlation
coefficients. It implies that it is difficult to detect obvious
double-peak AGNs with higher Eddington ratios. Using the
monochromatic luminosity at 5100\AA\ to trace the bolometric
luminosity, we find that the external illumination of the
accretion disk is needed to produce the observed strength of
H$\alpha$ emission line.

\end{abstract}

\section{INTRODUCTION}
Double-peaked broad Balmer emission profiles have been detected in
about 200 active galactic nuclei (AGNs) (Eracleous \& Halpern
1994, 2003; Strateva et al. 2003, 2006). A systematic survey of
110 sources (mostly broad-line radio loud AGNs at $z<0.4$) by
Eracleous \& Halpern (1994, 2003) suggested that about 20\%
sources show double-peaked broad Balmer lines. They found some
characteristics of double-peaked AGNs: large full width at half
maximum (FWHM) of H$\alpha$ line, ranging from about 4000 \kms \
up to 40000 \kms (e.g. Wang et al. 2005); about 50\% starlight
contribution in optical continuum around H$\alpha$; large
equivalent widths of low-ionization forbidden lines; large [O
I]/\OIII \ ratios. Strateva et al. (2003) found about 3\% of the
$z<0.332$ AGNs in the Sloan Digital Sky Survey (SDSS) are
double-peaked AGNs, the fraction  is smaller  than that in
radio-loud AGNs sample. Double-peaked lines have also been
detected in some low-ionization nuclear emission line regions
(LINERs), e.g. NGC1097, M81, NGC4450, NGC4203, NGC4579
(Storchi-Bergmann et al. 1993; Bower et al. 1996; Shields et al.
2000; Ho et al. 2000; Barth et al. 2001). It remains a matter as
debate of the origin of the double-peaked profiles.

There are mainly three models to interpret the origin of the
double-peaked profiles: the accretion disk (Chen \& Halpern 1989;
Eracleous \& Halpern 1994, 2003; Gezari et al. 2007); the
biconical outflow (Zheng et al. 1991; Abajas et al. 2006), and an
anisotropic illuminated BLRs (Goad \& Wanders 1996). More
recently, observational test and physical consideration
preferentially suggested that double-peaked profiles originate
from the accretion disk within a radius from a few hundreds $R_g$
to about thousands $R_g$ ($R_{g}=GM_{\rm bh}/c^2$, $M_{\rm bh}$ is
the SMBH mass) (Eracleous \& Halpern 1994, 2003; Eracleous et al.
1997; Strateva et al. 2003, 2006; Gezari, et al. 2007). In order
to interpret the sparsity of double-peaked AGNs, the origin of the
single-peaked lines from accretion disk have been discussed
(Eracleous \& Halpern 2003 and the references therein): larger out
radius of the line-emitting accretion disk; face-on accretion
disk; and the accretion disk wind. Very few of the double-peaked
high-ionization line profiles (e.g. CIV) is due to that these high
ionization lines are thought to arise in a wind, not in the disk.

The masses of central supermassive black holes (SMBHs) can provide
an important tool to understand the physics of double-peaked AGNs
if we reliably estimate them (e.g. Lewis \& Eracleous 2006; Lewis
2006). During the last decade, there is a striking progress on the
study of central supermassive black holes. The stellar and/or
gaseous dynamics is used to derive the SMBHs masses in nearby
inactive galaxies. However, for AGNs, this method is very
difficult because nuclei outshine their hosts. Fortunately, we can
use the broad emission lines from BLRs (e.g. H$\beta$, H$\alpha$,
Mg $\rm \sc II$, C$\rm \sc IV$) to estimate SMBH masses in AGNs by
the reverberation mapping method and the empirical size-luminosity
relation (Kaspi et al. 2000,2005; Vestergaard 2002; McLure \&
Jarvis 2002; Wu et al. 2004; Greene \& Ho 2006a). There is a
scaling factor with larger uncertainty, which is due to the
unknown geometry and dynamics of broad line regions, BLRs (e.g.
Krolik 2001; Collin et al. 2006). Nearby galaxies and AGNs seem to
follow the same tight $M_{\rm bh}-\sigma_{*}$ relation, where
$\sigma_{*}$ is the bulge velocity dispersion at eighth of the
effective radius of the galaxy (Nelson et al. 2001; Tremaine et
al. 2002; Greene \& Ho 2006a, 2006b), although it remains
controversial for narrow-line Seyfert 1 galaxies (NLS1s) (Mathur
et al. 2001; Bian \& Zhao 2004; Grupe \& Mathur 2004; Watson et
al. 2007). Lewis \& Eracleous (2006) derived the black hole masses
from $\sigma_*$ through fitting absorption lines of the Ca II
triplet ($\lambda \lambda8498, 8542, 8662 $) for 5 double-peaked
AGNs. Therefore the determination of the black hole mass from
independent method of reverberation mapping is an useful prober to
explore mysteries of the double peaked AGNs.

Significant stellar contribution in double-peaked AGNs makes the
measurement of $\sigma_*$  possible and reliable. Here we present
our results on the $\sigma_*$ measurements for the sample of 110
double-peaked AGNs from the Sloan Digital Sky Survey (SDSS)
(Strateva et al. 2003). In section 2, we introduce our working
sample selected from Strateva et al. (2003). Section 3 is data
analysis. We present the calculations of the SMBH mass and the
Eddington ratio, and discuss their errors in Section 4. Section 5
is contributed to the discuss of the BLRs in double-peaked AGNs.
Section 6 is the relation between the peak separation and
Mass/Eddington ratio. Section 7 is the Energy budget for
double-peaked AGNs. Our conclusion is given in Section 8. The last
section is our conclusion. All of the cosmological calculations in
this paper assume $H_{0}=71 \rm {~km ~s^ {-1}~Mpc^{-1}}$,
$\Omega_{M}=0.27$, and $\Omega_{\Lambda} = 0.73$.

\section{Sample}
Strateva et al. (2003) presented a sample of double-peaked AGNs
($z<0.332$) selected from SDSS. They used two steps to select
candidates: (1) they selected the unusual ones from the symmetric
lines using the spectral principal component analysis (PCA) (Hao
et al. 2003); and (2) they fitted the H$\alpha$ region with a
combination of several Gaussians, and only selected AGNs better
fitted by two Gaussians. From the profiles of the broad
components, several fitting parameters including the positions of
the red and blue peaks ($\lambda_{\rm red}$, $\lambda_{\rm
blue}$), the corresponding peak heights ($H_{\rm red}$, $H_{\rm
blue}$), FWHMs were given in their Table 3. They investigated  the
multi-wavelength properties of these double-peaked AGNs and
suggested that Eddington ratios could be large in these SDSS
double-peaked AGNs (Strateva et al. 2006).

SDSS spectra cover the wavelength range of 3800-9200 \AA\ with a
spectral resolution of $1800 < R < 2100$. The fiber diameter in
the SDSS spectroscopic survey is 3" on the sky. At the mean
redshift of 0.24 in the sample of Strateva et al. (2003), the
projected fiber aperture diameter is 13.2 kpc, typically
containing about 80\% of the total host galaxy light (Kauffmann \&
Heckman 2005). The stellar absorption features in these SDSS
spectra provide us the possibility to measure the stellar velocity
dispersion. We did not apply aperture corrections to the stellar
velocity dispersions because this effect can be omitted for
$z<0.3$ (Bernardi et al. 2003; Bian et al. 2006).

\section{Data analysis}
There are a number of objective and accurate methods to measure
$\sigma_{*}$, including two main different techniques, the
"Fourier-fitting" method (Sargent et al. 1977; Tonry \& Davis
1979), and the "direct-fitting" method (Rix \& White 1992; Greene
\& Ho 2006b and reference therein). With the development of
computing, the "direct-fitting" method become much more popular.
For SDSS spectra with significant stellar absorption features
(such as Ca H+K $\lambda \lambda$ 3969, 3934, Mg Ib$\lambda
\lambda$ 5167, 5173, 5184 triplet, and Ca II$\lambda \lambda$
8498, 8542, 8662 triplet, etc.), $\sigma_{*}$ can be measured by
matching the observed spectra with the combination of different
stellar template spectra broadened by a Gaussian kernel (e.g.
Kauffmann et al. 2003; Cid Fernandes et al. 2004; Vanden Berk et
al 2006; Greene \& Ho 2006b; Zhou et al. 2006; Bian et al. 2006).
The SMBH mass can then be estimated from the $M_{\rm
bh}-\sigma_{*}$ relation if $\sigma_{*}$ can be accurately
measured from the spectrum of  AGN host galaxy.

Here we briefly introduce the method to measure $\sigma_{*}$.
Adopting the stellar library from Bruzual \& Charlot (2003), we
used the stellar population synthesis code, STARLIGHT, (Cid
Fernandes et al. 2004, 2005; Bian et al. 2006) to model the
observed spectrum $O_\lambda$. It models the spectrum $M_\lambda$
by the linear combination of simple stellar populations (SSP). The
model spectrum $M_\lambda$ is:
\begin{equation}
M_\lambda(x,M_{\lambda_0},A_{\rm V},v_*,\sigma_*) = M_{\lambda_0}
   \left[
   \sum_{j=1}^{N_*} x_{\rm j} b_{\rm j,\lambda} r_\lambda
   \right]
   \otimes G(v_*,\sigma_*)
\end{equation}
where $b_{\rm j,\lambda} \equiv L_\lambda^{\rm SSP}(t_{\rm
j},Z_{\rm j}) / L_{\lambda_0}^{\rm SSP}(t_{\rm j},Z_{\rm j})$ is
the $j^{\rm th}$ template normalized at $\lambda_0$, $x$ is the
population vector, $M_{\lambda_0}$ is the synthetic flux at the
normalization wavelength, $r_{\lambda}$ is the reddening term,
$A_{\rm V}$ is the $V$-band extinction, and $G(v_*,\sigma_*)$ is
the line-of-sight stellar velocity distribution, modelled as a
Gaussian centered at velocity $v_*$ and broadened by $\sigma_*$.
The best fit is reached by minimizing $\chi^2$,

\begin{equation}
\chi^2(x,M_{\lambda_0},A_{\rm V},v_*,\sigma_*) =
   \sum_{\lambda=1}^{N_\lambda}
   \left[
   \left(O_\lambda - M_\lambda \right) w_\lambda
   \right]^2
\end{equation}
where the weighted spectrum $w_\lambda$ is defined as the inverse
of the noise in observed spectra. We can obtained some parameters
including $\sigma_*$ by modelling the observed spectrum, which are
important for our understanding the stellar populations in AGNs
host galaxies.

We download 110 double-peaked spectra from SDSS
DR5\footnote{http://cas.sdss.org/dr5/en/tools/search/} in the
sample of Strateva et al. (2003), as well as the extinction values
at $g$ band. We perform the Galactic extinction correction to
observed spectra by using the extinction law of Cardelli, Clayton
\& Mathis (1989). Then, the rest-frame spectra with errors and
masks are obtained by use of the redshift given in their SDSS FITS
headers. We normalize the spectra at 4020\AA\ and the
signal-to-noise ratio (S/N) is measured between 4010 and 4060 \AA.
An additional power-law component is used to represent the AGNs
continuum emission. We check visually our modelled spectra sorted
by the equivalent width (EW) of CaII K $\lambda 3934$ line. It is
found that the fitting goodness depends on the S/N ($\gtrsim5$),
the absorption lines equivalent widths (EW (CaII K) $\gtrsim$ 1.5
\AA), and the contribution of stellar lights (see also Zhou et al.
2006). At last, 54 double-peaked AGNs are selected, which are well
fitted to derive reliable $\sigma_*$ from their significant
stellar absorption. Using the two sample Kolmogorov-Smirnov (K-S)
test, {\it kolmov} task in IRAF, the distributions of the
luminosity at 5100\AA\ in the total sample of Strateva et al.
(2003) and our sub-sample are drawn from the same parent
population with the probability of 53\%. Therefore, this
sub-sample is representative of the total sample of Strateva et
al. (2003) with respect to the luminosity at 5100\AA. Fig. 1 shows
a fitting example for SDSS J082113.71+350305.02. The final results
are presented in Table 1, arranging in order of increasing right
ascension.

\begin{figure*}
\begin{center}
\includegraphics[width=10cm,height=13cm,angle=-90]{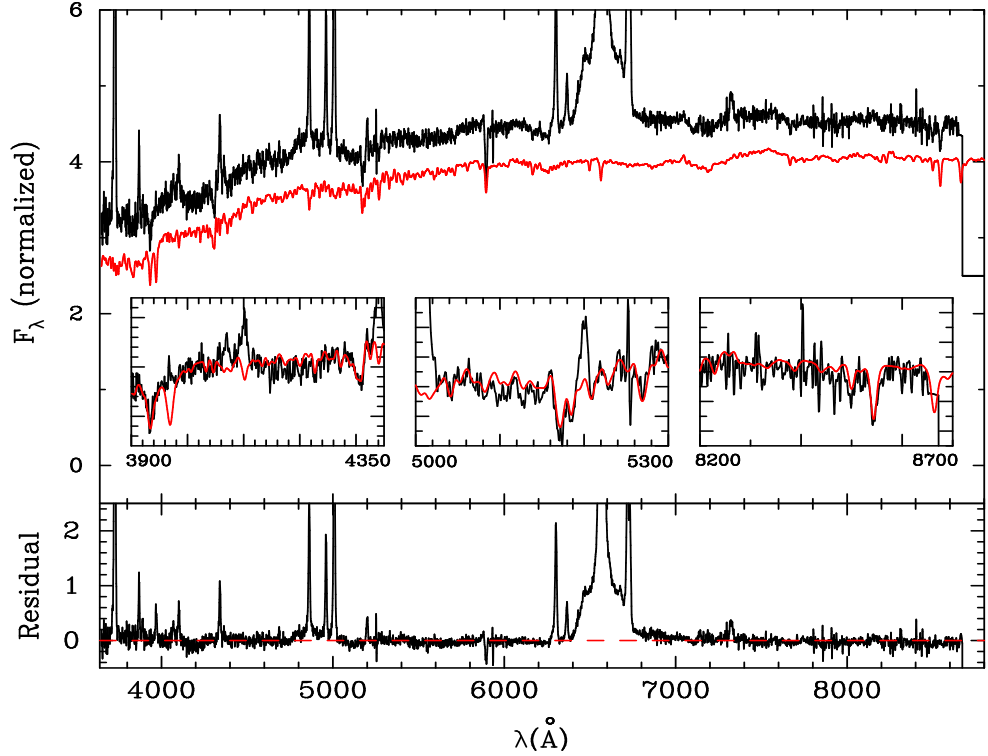}
\caption{Sample fit of the synthetic population model for SDSS
J081700.40+34556.34. The top panel is the logarithm of the
observed spectrum (top black curve) and the synthetic spectra
(bottom red curve)(shifted down for clarity). In this top panel,
we also show the region around Ca H+K $\lambda \lambda$ 3969, 3934
and G-band (left); the region around Mg Ib$\lambda \lambda$ 5167,
5173, 5184 triplet, Fe 5270\AA\ (middle); the region around Ca
II$\lambda \lambda$ 8498, 8542, 8662 triplet (right). The residual
spectrum is shown in the bottom panel. }
\end{center}
\end{figure*}

Using Interactive Data Language (IDL), we also carefully model the
profiles of \OIII\ and H$\beta$ lines. Since the double-peaked
profile of the H$\beta$ line and the asymmetric profiles of
\OIII$\lambda\lambda$4959, 5007 lines, seven-gaussian profiles are
used to model these lines carefully, two broad and one narrow
components for H$\beta$ plus two sets of one broad and one narrow
components for \OIII$\lambda \lambda 4959, 5007$. We take the same
linewidth for each component of \OIII$\lambda \lambda 4959, 5007$,
fix the flux ratio of \OIII$\lambda$4959 to \OIII$\lambda$5007 to
be 1:3, and set their wavelength separation to the laboratory
value. And we also set the wavelength separation between the
narrow component of H$\beta$ and the narrow \OIII$\lambda$5007 to
the laboratory value. We do the lines fitting for these 54
stellar-light subtracted spectra. For spectra without obvious
stellar features, we do the emission lines fitting for the
extinction-corrected rest-frame SDSS spectra without stellar-light
subtraction. At last we obtain the gaseous velocity dispersion,
$\sigma_g$, from the core \OIII$\lambda 5007$ line, the total flux
of \OIII$\lambda 5007$ line and the monochromatic flux at 5100\AA.
Our emission-line profile fitting for SDSS J022014.57-072859.30 is
shown in Fig. 2.

\begin{figure*}
\begin{center}
\includegraphics[width=13cm,height=8cm]{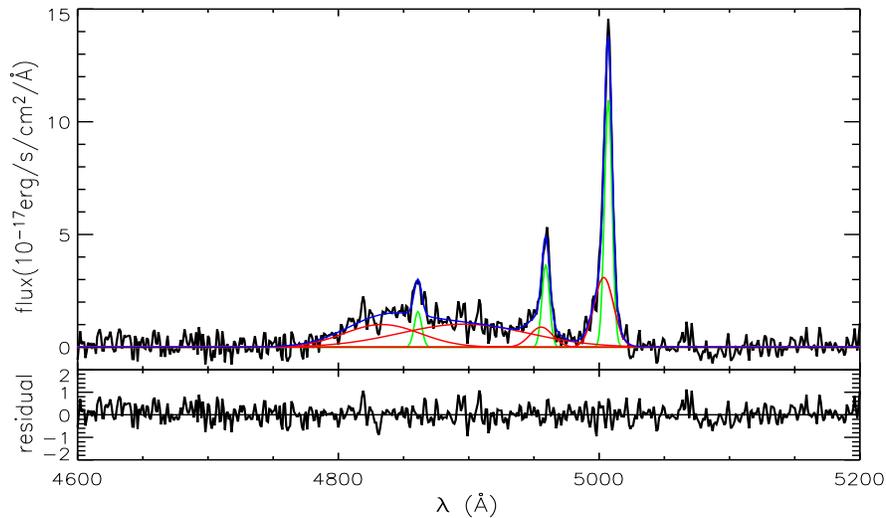}
\caption{Sample multi-component fitting of the
\OIII$\lambda\lambda$4959, 5007 lines for SDSS
J022014.57-072859.30: composite profile(blue curves); narrow
components (green curves); broad components (red curves); the
residue (bottom panel).}
\end{center}
\end{figure*}

\section{The Masses and the Eddington ratios}
\subsection{The stellar velocity dispersions}
\begin{figure}
\begin{center}
\includegraphics[width=9cm]{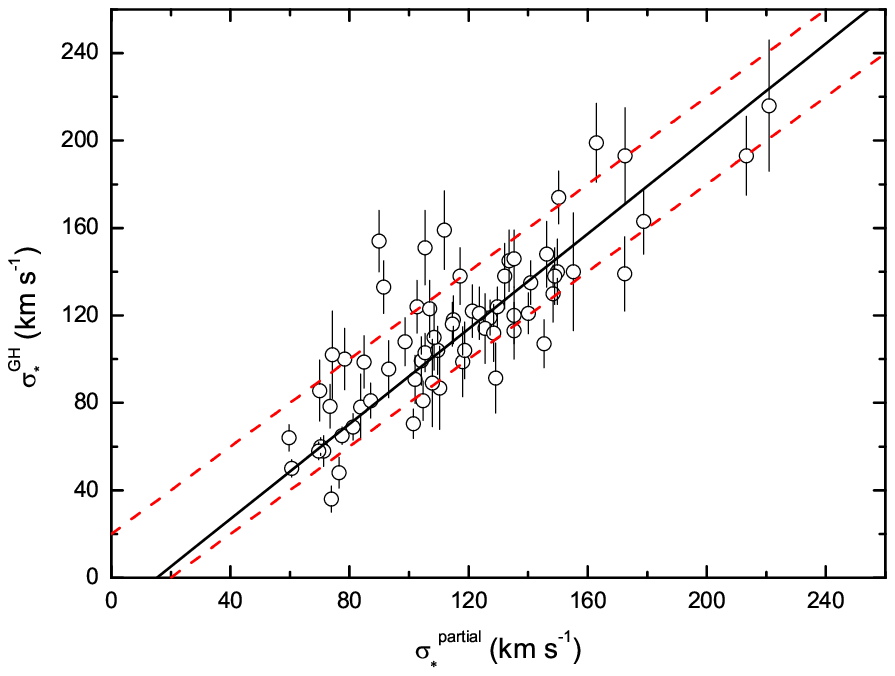}
\caption{Comparison of our $\sigma_*$ from the partial SDSS
spectra with that for a sample of the local AGNs presented by
Greene \& Ho (2006a). The solid line denotes our best linear fit,
the red dash lines are $y=x\pm 20$ \kms. Considering the error bar
of $\sigma_*$ presented by Greeene \& Ho (2006a), most of local
AGNs are located between two red dash lines.}
\end{center}
\end{figure}

In the synthesis, we focus on the strongest stellar absorption
features, such as CaII K, G-band, and Ca II$\lambda \lambda$ 8498,
8542, 8662 triplet, which are less affected by emission lines. We
put twice more weight for these features during the stellar population synthesis.
After correcting the template and the SDSS instrumental resolution, we obtain the value of
$\sigma_*$ through the direct-fitting method.

Cid Fernandes et al. (2005) applied the same synthesis method to a
larger sample of 50,362 normal galaxies from the SDSS Data Release
2 (DR2). Their $\sigma_{*}$ is consistent very well with that of
the MPA/JHU group (Kauffmann et al. 2003), the median of the
difference is just 9 \kms. They also gave the $\sigma_*$
uncertainty based on the S/N (see Table 1 in Cid Fernandes et al.
2005). Typically, the $\sigma_*$ uncertainty based on S/N at
4020\AA\ is about: 24 \kms at S/N=5; 12 \kms at S/N=10; 8 \kms at
S/N=15. In order to use the typical errors suggested by Cid
Fernandes et al. (2005), we calculate the S/N and the starlight
fraction at 4020\AA. The S/N is the mean flux divided by the root
mean square (RMS) of the flux in the range 4010 to 4060 \AA. We
also performed the method of Cid Fernandes et al. (2005) to
compute the S/N by using the SDSS error spectrum, the results are
almost the same. Considering the contribution from featureless
continuum (FC, represented by a power law), we use an effective
S/N to show the typical error of $\sigma_*$, where the effective
S/N is roughly the S/N multiplied by the stellar fraction. We use
the featureless continuum fraction as the up limit of nuclei
contribution, because the featureless continuum can be attributed
either by a young dusty starburst, by an AGN, or by these two
combination (Cid Fernandes et al. 2004). Therefore, the effective
S/N is the low limit. In our sample, the effective mean spectral
S/N at 4020\AA\ for these objects are 7.2 (see Table 1). Thus the
typical uncertainty in $\sigma_{*}$ should be around 20 \kms. For
13 objects with effective S/N less than 5, their $\sigma_*$ in
Table 1 are preceded by colons.

Here, we also apply this synthesis method to a sample of the local
AGNs presented by Greene \& Ho (2006a). Greene \& Ho (2006a;
2006b) performed a research on the systematic bias of $\sigma_{*}$
derived from the regions around CaII triplet, MgIb triplet, and
CaII H+K, respectively (Barth et al. 2002). They argued that the
CaII triplet provide the most reliable measurements of
$\sigma_{*}$ and there is a systematic offset between $\sigma_{*}$
from CaII K line and that from other spectral regions. We use our
synthesis method to their sample in two manners: one is using
whole spectrum, the other is just using partial spectrum between
3200\AA\ and 7500\AA \ at the rest frame. We put twice more weight
for the strongest absorption features of Ca H+K $\lambda \lambda$
3969, 3934, G-band, and Ca II$\lambda \lambda$ 8498, 8542, 8662
triplet.  We find that the values of $\sigma_*$ in these two
manners are similar by performing our synthesis method. By using
the least-square regression, the best fit between the $\sigma_*$
from these two manners ($\sigma^{\rm whole}_*$ and
$\sigma_{*}^{\rm partial}$, respectively) is: $\sigma^{\rm
whole}_* =(5.49\pm3.29)+(0.95\pm0.03)\sigma_{*}^{\rm partial}$.
The spearman coefficient $R$ is 0.97, with a probability of
$p_{\rm null} < 10^{-4}$ for rejecting the null hypothesis of no
correlation.

Because the SDSS spectral coverage is from 3800\AA\ to 9200\AA,
most of double-peaked AGNs with redshift larger than 0.083 will
not cover the range of CaII triplet. For those sources with
redshift larger than 0.083, we used the above formula to obtain
the corrected velocity dispersion, i.e. $\sigma^{\rm c}_*
=(5.49\pm3.29) +(0.95\pm0.03)\times \sigma_{*}$. The corrected
velocity dispersion is listed in Col. (8) in Table 1. And we find
that the corrected velocity dispersion is almost the same to the
uncorrected one. The largest difference is about 10 \kms, which is
less than the typical error of 20\kms.

Based on the SDSS instrumental resolution, we take 60 \kms \ as a
lower limit of $\sigma_*$ (e.g. Bernardi 2003).  Only for two
objects, SDSS J133433.24 -013825.41 and SDSS J214555.03
+121034.17, the measured $\sigma_*$ are below/near the lower limit
(51 \kms and 62 \kms, respectively). These two objects are
excluded from further analysis.

The $\sigma_*$ value from the partial SDSS spectra is also used to
compare our result with Greene \& Ho (2006a), who fitted the
region around CaII triplet directly. In Fig. 3, we compared our
$\sigma_*$ with theirs. We found that the agreement is quite good,
the $\sigma_*$ difference ($\sigma_*^{\rm GH} -\sigma_*^{\rm
partial}$) distribution is -2.1 \kms \ with a standard deviation
(SD) of 22.7 \kms. By using the least-square regression
considering the errors of $\sigma_*^{\rm GH}$, the best fit
between $\sigma^{\rm GH}_*$ and $\sigma_{*}^{\rm partial}$ is:
$\sigma^{\rm GH}_* =(-16.66\pm3.56)+(1.09\pm0.04)\sigma_{*}^{\rm
partial}$ (solid line in Fig. 3). The spearman coefficient $R$ is
0.86, the standard deviation is 2.11, with $p_{\rm null} <
10^{-4}$.

\subsection{The Results}
For these double-peaked AGNs with reliable $\sigma^{c}_*$, we use
the $M_{\rm bh}-\sigma_{*}$ relation to derive the SMBH mass (e.g.
Tremaine et al. 2002 and reference therein),
\begin{equation}
M_{\rm bh} (\sigma_*^c) = 10^{8.13}\left(\frac{\sigma_{*}^{c}}
                           {200~ \rm km~s^{-1}}\right)^{4.02} ~~\Msun .
\end{equation}
We calculate the Eddington ratio, e.g., the ratio of the
bolometric luminosity ($L_{\rm bol}$) to the Eddington luminosity
($L_{\rm Edd}$), where $L_{\rm Edd}=1.26 \times 10^{38} (M_{\rm
bh}/\Msun) \ \ergs$. We use the monochromatic luminosity at
5100\AA\ ($\lambda L_{\lambda}$ at 5100\AA) to estimate the
bolometric luminosity, $L_{\rm bol}=c_{\rm B} \lambda
L_{\lambda}$(5100\AA), where $c_{\rm B} = 9$ (Kaspi et al. 2000).
These results are presented in Table 1.

For the typical uncertainty of 20 \kms \ for $\sigma_{*}=200$
\kms, the error of $\log~ \sigma_{*}$ would be about 0.05 dex,
corresponding to 0.2 dex for $\log M_{\rm bh}$. The error of $\log
M_{\rm bh}$ is about 0.4 considering the error of 0.3 dex form the
$M_{\rm bh}-\sigma_{*}$ relation (Tremaine et al. 2002). Richards
et al. (2006) suggested a bolometric correction factor of $10.3\pm
2.1$ at 5100 \AA. Therefore, the final Eddington ratio, $L_{\rm
bol}/L_{\rm Edd}$, has a large uncertainty, about 0.5 dex.

\begin{figure}
\begin{center}
\includegraphics[width=9cm]{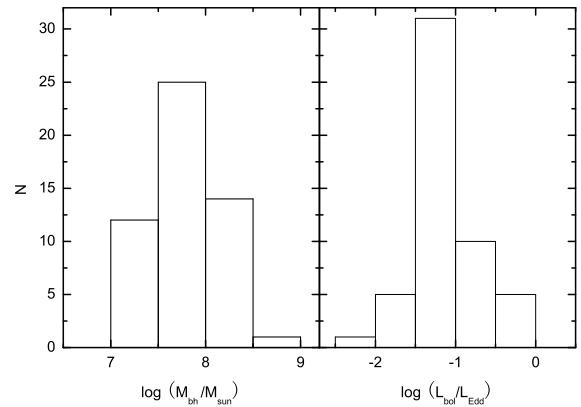}
\caption{The histograms of the black hole masses (left) and the
Eddington ratios (right) for these 52 double-peaked AGNs with
reliable $\sigma^{c}_{*}$.}
\end{center}
\end{figure}

In Fig. 4, we show the histograms of the black hole masses and the
accretion ratios for these 52 double-peaked AGNs. The black hole
masses range from $1.0\times 10^{7}$ to $5.5\times 10^{8}$ $\Msun$
with a mean value of $\log M_{\rm bh}/\Msun=7.76 \pm 0.37$. Using
the H$\alpha$ FWHM and the 5100\AA\ monochromatic luminosity, Wu
\& Liu (2004) estimated the SMBH masses and the Eddington ratios
for an assembled double-peaked AGNs sample. Lewis \& Eracleous
(2006) noted that the BH masses from the H$\alpha$ FWHM are not
completely consistent with those from the stellar velocity
dispersion. Our BH masses derived from $\sigma^c_*$ are indeed
smaller than those from the H$\alpha$ FWHM (Wu \& Liu 2004) by
about an order of magnitude. This will lead to our $L_{\rm
bol}/L_{\rm Edd}$ larger than that from the H$\alpha$ FWHM by
almost an order of magnitude.

\begin{figure}
\begin{center}
\includegraphics[width=9cm,height=7cm]{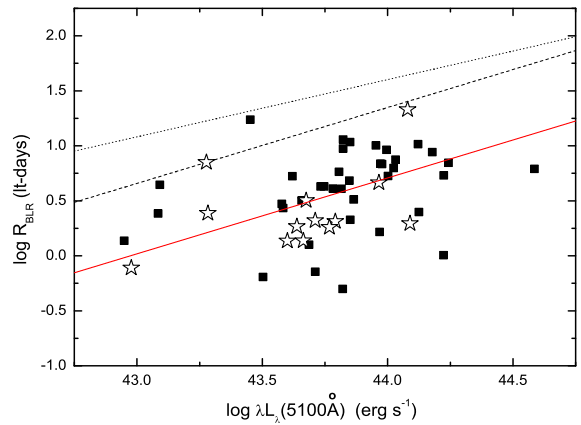}
\caption{The BLRs sizes versus the monochromatic luminosity at
5100\AA\ (where $f$=0.75 for random BLRs orbits). Objects with
effective S/N in 4020 \AA\  less then 5 is denoted as open stars.
The black dash line is the empirical size-luminosity relation,
$R_{\rm BLR}=22.3\times [\lambda
L_{\lambda}$(5100\AA)$/10^{44}~\rm erg s^{-1}]^{0.69} ~\rm
lt-days$, found by Kaspi et al. (2005). The black dot line is the
empirical size-luminosity relation,$R_{\rm BLR}=40\times (\lambda
L_{\lambda}$(5100\AA)$/10^{44}~\rm erg s^{-1})^{0.52} ~\rm
lt-days$, found by Bentz et al. (2006). The red solid line is our
best fit with the fixed slope of 0.69 ($R$=0.29), lower by -0.63
dex respect to dash line.}
\end{center}
\end{figure}

The Eddington ratio has a distribution with mean and the standard
deviation of $\log (L_{\rm bol}/L_{\rm Edd})$ is $-1.13 \pm 0.38$.
It is suggested that the accretion disk in double-peaked AGNs is
in the mode of Advection Dominated Accretion Flow (ADAF) (e.g.
Eracleous \& Halpern 2003). When the Eddington ratio is below than
the critical one $L_{\rm Bol}/L_{\rm Edd}\sim
0.0028\alpha_{0.1}^{2}$ (Mahadevan 1997), the ADAF appears, where
$\alpha_{0.1}=\alpha/0.1$ is viscous coefficient. For our SDSS
sample, all objects have Eddington ratios larger than this
critical value of 0.0028. The present results from Fig. 4 clearly
show that these double peaked AGNs have accretion disks in the
standard regime.

The black hole masses can be independently tested by the relation
between the black holes and bulges. The Appendix gives details of
the test. We find the black hole masses are consistent from
$M_{\rm bh}-M_{\rm bulge}$ and $M_{\rm bh}-\sigma_*$ relations.

\section{BLRs in double-peaked AGNs}

\subsection{The size of BLRs}

We also calculate the BLRs sizes for these 52 double-peaked AGNs
using the SMBHs masses derived from $\sigma_*^c$ and the H$\alpha$
FWHM. We firstly transform the H$\alpha$ FWHM to the H$\beta$ FWHM
by (Greene \& Ho 2005b):
\begin{equation}
\rm FWHM_{\rm H \beta} = (1.07 \pm 0.07) \times 10^3 \left
(\frac{\rm FWHM_{\rm H \alpha}}{10^3~ \rm km s^{-1}}
\right)^{(1.03 \pm 0.03)}~{\rm km s^{-1}}.
\end{equation}

\noindent From the SMBH masses derived from the velocity
dispersions, we can calculate the BLRs sizes:
\begin{equation}
R_{\rm BLR} = \frac{M_{\rm bh}(\sigma_*^c)\times 5.123}{f \times
\rm FWHM_{ H \beta}^2} ~~~~~~~~~~\rm lt-days,
\end{equation}
where $f$ is the scaling factor related to the kinematics and
geometry of the BLRs, defined by $M_{\rm bh}=f\frac{R_{\rm BLR}
V_{\rm p}^2}{G}$, $V_{\rm p}$ is the Keplerian velocity in disk
plane. For random orietation of BLR cloud Keplerian orbits, $f$ is
0.75.

In Fig. 5, we plot the BLRs sizes versus the monochromatic
luminosity at 5100\AA\ (assuming $f$=0.75 for random BLRs orbits).
Almost all objects are located below the empirical size-luminosity
relation (black dash line) found by Kaspi et al. (2005). The
correlation between the BLRs sizes and the monochromatic
luminosity at 5100\AA\ is not too strong. The best fit is shown as
red solid line in Fig. 5 as $R_{\rm BLR}=5.1\left[\lambda
L_{\lambda}(5100\AA)/10^{44}~{\rm erg s^{-1}}\right]^{0.69}$
lt-days ($R=0.34$, $P_{\rm null}=0.20$), lower by -0.64 dex
respecting to the empirical relation found by Kaspi et al. (2005).
When excluding objects with effective S/N in 4020 \AA\ less then 5
(open stars in Fig. 5), for fixed slope of 0.69, the best fit
gives almost the same line but with a smaller $R$ of $0.28$.

If we used $f$=0.52 (Table 2 in Collin et al. 2006), $\log R$ will
increased by 0.16 dex, the BLRs sizes of double-peaked AGNs  are
still deviated from the empirical relation found by Kaspi et al.
(2005) (about 0.48 dex). FWHMs of the double-peaked AGNs are about
twice as broad as other AGN of similar luminosity (Eracleous \&
Halpern 1994, 2003; Strateva et al. 2003). If the double-peaked
AGNs are not systematically more massive than other AGNs, equation
5 suggested that BLRs radii can be smaller than for a similarly
massive "normal" AGNs by a factor of 4.

We have to point out that here we are not really testing the
$R_{\rm BLR}-L$ relation in double peaked AGNs, but the comparison
with $R_{\rm BLR}-L$ relation allows us to derive the factor $f$.

\subsection{ The factor $f$, BLR inclinations and Non-virial BLRs}
If we use the empirical $R_{\rm BLR}-L$ relation (Kaspi et al.
2005) to derive the $R_{\rm BLR}$ and the SMBH masses, $M_{\rm
bh}(\sigma_*^c)$, to do the calibration of the factor $f$ (Onken
et al. 2004), we find that the distribution of $f$ is 0.179 with a
standard deviation of 0.171. It is not consistent with the value
of $\langle f\rangle=5.5/4$ presented by Onken et al. (2004) and
it is about 1/3 of $0.52\pm 0.13$ suggested by Collin et al.
(2006). Our results suggest that the empirical relation between
the BLRs sizes and the luminosity at 5100\AA\ does not hold for
double-peaked AGNs (see also Zhang et al. 2007), otherwise the
calibration factor $f$ should be as low as 0.179. An important
consequence of the breakdown in the size-luminosity relationship
is that using mass estimation methods based on the size-luminosity
relationship and the calibration factor of 0.75 can lead to an
order of magnitude over-estimate of the SMBH mass (Wu \& Liu
2004).

If BLRs are disk-like with an inclination of $\theta$, the
relation between the H$\beta$ FWHM and the Keplerian disk plane
velocity, $v_{\rm p}$, is given by (Wills \& Browne 1986)
\begin{equation}
{\rm FWHM_{\rm H \beta}}=2\left(V_{\rm r}^{2}+V_{\rm p}^{2}
\sin^{2}\theta\right)^{1/2},
\end{equation}
where $V_{\rm r}$ is the random isotropic component. We may derive
the scaling factor as $f=1/[4(V_{\rm r}/V_{\rm p})^2+ 4\rm sin^2
\theta]$. Ignoring $V_{\rm r}$, $f=1/(\rm 2sin\theta)^2$, and the
minimum of $f$ is 0.25. For only ten object with $f>0.25$, we can
derive the inclination of $\theta$ by the above formula. The mean
inclination is 56 degrees. It is suggested that double-peaked AGNs
are not preferentially edge-on (Eracleous \& Halpern 1994, 2003;
Strateva et al. 2003), most have an inclination of less than 50
degrees. If all these objects were nearly edge-on, the obscuring
torus would prevent us from seeing the broad lines. Collin et al.
(2006) suggested that objects with large FWHMs, inclination
effects are not really important. In these 52 double-peaked AGNs,
42 objects have smaller $f$ less than 0.25. $V_{\rm r}$ can't be
omitted, we need to consider the random isotropic component,
implying the non-virial dynamics of BLRs in double-peaked AGNs. If
we use $f$ to trace the non-virial effect, smaller $f$ means
strong non-virial effect. We find $f$ have strong correlations
with $\log M_{\rm bh}$ and $\log(L_{\rm bol}/L_{\rm Edd})$ (see
Fig. 6). Using the least-square regression, we derive the
correlations are:
\begin{equation}
\log f =(-1.63 \pm 0.12)-(0.65\pm 0.10) \log(L_{\rm bol}/L_{\rm
Edd}),
\end{equation}
where $R=-0.69$, $p_{\rm null} < 10^{-4}$; and
\begin{equation}
\log f =(-4.31 \pm 0.97)+(0.44\pm 0.13) \log M_{\rm bh}/\Msun,
\end{equation}
where $R=0.45$ with $p_{\rm null} =9.6 \times 10^{-4}$. When we
exclude objects with effective S/N in 4020 \AA\ less then 5 (open
stars in Fig. 6), the best fits for $f$ relations between SMBH
mass and the Eddington ratio give a little larger $R$ ($-0.75,
0.52$, respectively). We should note that the reason of these
strong correlations are that $f$ is derived from $M_{\rm bh}$,
$R_{\rm BLR}$ [from $\lambda L_{\lambda} (5100\AA)$ ], and $\rm
FWHM_{H \beta}$. These strong correlations suggest that objects
with larger Eddington ratios have strong non-virial effect.

\begin{figure}
\begin{center}
\includegraphics[width=9cm]{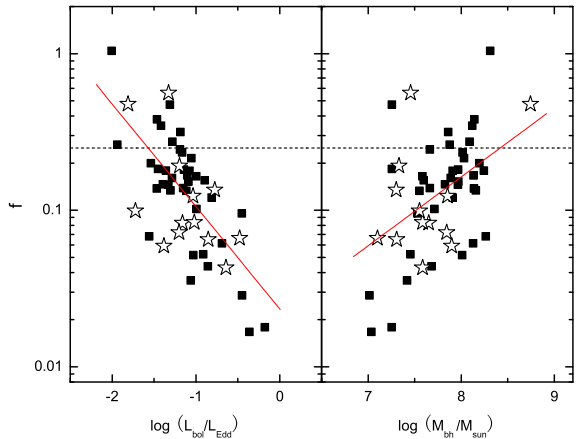}
\caption{The scaling factor $f$ versus the Eddington ratio (left
panel) and the SMBH masses (right panel). Objects with effective
S/N in 4020 \AA\  less then 5 is denoted as open stars. The dash
lines are $f=0.25$. The correlation coefficient for the relation
between the peak separation and the Eddington ratio is $R=-0.69$
and that between the separation and the black hole mass is
$R=0.45$.}
\end{center}
\end{figure}

3C 390.3 is the only double peaked object with the direct
measurements of the time delay and the host stellar velocity
dispersion. And it locates in the empirical relation between BLRs
sizes and the luminosity (e.g. Kaspi et al. 2005). By $\sigma_*$
of 240 \kms (Onken et al. 2004), we find that $f$ is about 1.3,
consistent with the assumption of random BLRs orbits in 3C390.3.
Its small Eddington ratio (Lewis \& Eracleous 2006), $(2-4)\times
10 ^{-2}$, is also consistent with the virial BLRs dynamics (See
Fig. 6).

\section{Peak separation, BH Mass and Eddington ratio}
\begin{figure}
\begin{center}
\includegraphics[width=9cm]{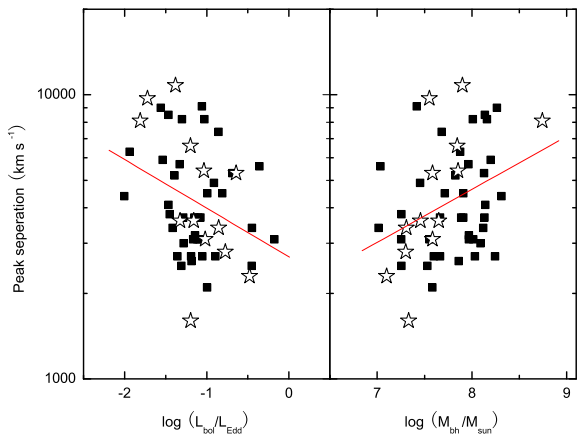}
\caption{The peak separation versus the Eddington ratio (left
panel) and the SMBH masses (right panel). Objects with effective
S/N in 4020 \AA\ less then 5 is denoted as open stars. The
correlation coefficient for the relation between the peak
separation and the Eddington ratio is $R=-0.33$, which is almost
the same for that between the separation and the black hole mass
($R=0.35$).}
\end{center}
\end{figure}

For double-peaked AGNs, the peak separation, $\Delta v =
\lambda_{\rm red} -\lambda_{\rm blue} $, has a large variance
ranging from about 1000 \kms\ to  about 10000 \kms\ (Table 3 in
Strateva et al. 2003). In Fig. 7, we show the relations between
$\log \Delta \lambda$ and $\log(L_{\rm bol}/L_{\rm Edd})$, $\log
M_{\rm bh}$. Using the least-square regression, we derive the
correlation between $\log \Delta v$ and $\log(L_{\rm bol}/L_{\rm
Edd})$ to be:
\begin{equation}
\log \Delta v =(3.43 \pm 0.08)-(0.17\pm 0.06) \log(L_{\rm bol}/L_{\rm Edd}),
\end{equation}
where $R=-0.33$, $p_{\rm null} = 0.017$ (see Fig. 7). For the
relation between $\log \Delta \lambda$ and $\log M_{\rm bh}$,
\begin{equation}
\log \Delta v =(2.17 \pm 0.55)+(0.11\pm 0.08) \log (M_{\rm
bh}/\Msun),
\end{equation}
where $R$ is 0.35 with $p_{\rm null} = 0.011$. When excluding
objects with effective S/N in 4020 \AA\ less then 5 (open stars in
Fig. 7), the best fits for $\log \Delta v$ relations between SMBH
mass and the Eddington ratio give smaller $R$ ($-0.22, 0.21$,
respectively). We also do the multiple regression for the
dependence of $\Delta v$ on $\log(L_{\rm bol}/L_{\rm Edd})$ and
$\log M_{\rm bh}$,
\begin{eqnarray}
\log \Delta v =(2.54 \pm 0.64)+(0.13\pm 0.09) \log (M_{\rm
bh}/\Msun) \\ \nonumber - (0.10\pm 0.09)\log(L_{\rm bol}/L_{\rm
Edd}),
\end{eqnarray}
where the $R-$Square correlation coefficient is 0.15.

Wu \& Liu (2004) also studied this correlation and obtained an
apparent stronger correlation ($R=0.84$). Their derived strong
correlation is mainly due to the very strong correlation between
the separation and the H$\alpha$ FWHM ($R=0.88$). When the
H$\alpha$ FWHM is fixed, they found that the partial correlation
coefficient is only 0.05. In this paper, we use the stellar
velocity dispersion to estimate the black hole masses, and we also
find a mild correlation between $\Delta v$ and $L_{\rm bol}/L_{\rm
Edd}$, which implies that the peak separation would be smaller for
AGNs with higher Eddington ratios. It provides clues to why
previous double-peaked AGNs have lower Eddington ratios.

There is increasing evidence for a disk geometry of the BLR (see a
review of Laor 2007). If we assume that peak separation is created
by the doppler shift of the movement of a thin annulus and the
annulus radius corresponds the location where the self-gravitation
domiantes (Bian \& Zhao 2002), we have the radius $R_{\rm
SG}/R_g=3586m_8^{-2/9}\dot{m}^{4/9}\alpha^{2/9}$ (their eq. 18 in
Laor \& Netzer 1989), where $R_g=1.5\times 10^{13}m_8$~cm,
$m_8=M_{\rm bh}/10^8\Msun$, $\dot{m}$ is the Eddington ratio and
$\alpha$ is the viscosity. The maximum separation of the double
peaks in units of \AA\ under the edge-on orientation to an
observer is given
\begin{equation}
\Delta \lambda=\lambda_0\left(\frac{R_{\rm g}}{R_{\rm SG}}\right)^{1/2}
              =84.9~\alpha_{0.1}^{-1/9} \dot{m}^{-2/9}m_8^{1/9}~\AA,
\end{equation}
where we use the peak separation in term of separation velocity,
\begin{equation}
\Delta v= 7.76 \times 10^3 ~\alpha_{0.1}^{-1/9}
\dot{m}^{-2/9}m_8^{1/9}~ \kms.
\end{equation}
where $\alpha_{0.1}=\alpha/0.1$ and H$\alpha$ wavelength $\lambda_0=6563$\AA.

Considering the uncertainties of the fitting results for the peak
separation correlations with SMBH mass and the Eddington ratio,
the slopes are consistent with the simple theoretical expectation.
We have to stress that equation (11) is for the maximum separation
(edge-on orientation) when the disk structure is given. More sophisticated
model is needed for explanations of the dependence of peak separations and
Eddington ratios. The
scatters in Fig 7 may be caused by different orientation of the
BLR in the sample.

We also find no significant correlations between the ratio of the
red peak height to the blue height and the Eddington ratio, the
black hole mass, the peak separation. Since the Keplerian velocity
is much below the light speed, the Doppler boosting effects could
be hidden by  complex situations of the BLR.

\begin{figure*}
\begin{center}
\includegraphics[width=7cm,height=13cm,angle=-90]{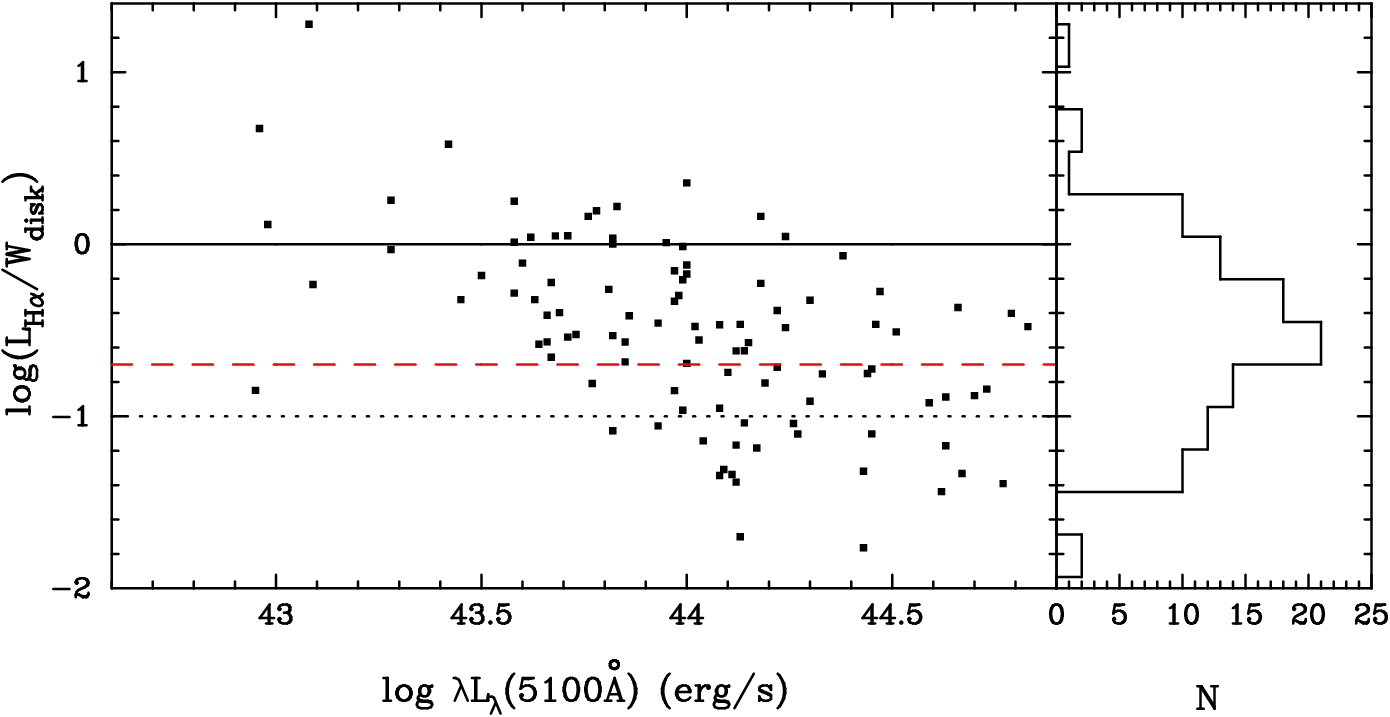}
\caption{The ratio of the H$\alpha$ line luminosity ($L_{\rm
H\alpha}$) to the line-emitting power ($W_{\rm disk}$) versus the
monochromatic luminosity at 5100\AA\ ($\log \lambda
L_{\lambda}$(5100 \AA)). The solid line is $L_{\rm H\alpha}=W_{\rm
disk}$, the dash line is $L_{\rm H\alpha}=0.2 W_{\rm disk}$, and
the dot line is $L_{\rm H\alpha}=0.1 W_{\rm disk}$. We display the
histogram of the distribution of $L_{\rm H \alpha}/W_{\rm disk}$
(right).}
\end{center}
\end{figure*}

\section{Energy budget}

Here we discuss the energy budget of the line-emitting accretion
disk. Based on the standard accretion disk model, the disk
radiation as a function of radius $\xi$ is (Chen et al. 1989):
\begin{equation}
F(\xi)=1\times 10^{12}
m_{8}^{-2}\dot{M}_{24}\xi_{100}^{-3}[1-(6/\xi)^{1/2}]~ \rm ergs~
cm^{-2}~s^{-1}
\end{equation}
where $m_8=M_{\rm bh}/10^8\Msun$, $\dot{M}_{24}$ is the accretion
rate in units of $10^{24} ~g~ s^{-1}$, and the dimensionless
radius $\xi_{100}$ is $\xi/(100R_{g})$. Using $\dot{M}=L_{\rm
bol}/(\zeta c^2)$ where the efficiency $\zeta$ is 0.1, the
gravitational power output of the line-emitting disk annulus
between $\xi_1$ and $\xi_2$ is (Eracleous \& Halpern 1994):
\begin{eqnarray}
W_{\rm disk}(\xi_1,\xi_2)=7.7\times L_{\rm bol}~~~~~~~~~~~~~~~~~~~\nonumber \\
 \left
[\frac{1}{\xi_{1}} \left (1-{\sqrt \frac{8}{3\xi_{1}}}\right
)-\frac{1}{\xi_2}\left (1-\sqrt \frac{8}{3\xi_2}\right)\right ]
{\rm ergs ~s^{-1}},
\end{eqnarray}
where $L_{\rm bol}$ is in units of $\ergs$.  It is noted that
$W_{\rm disk}$ is independent of the black hole mass, when we use
the typical radius in units of $R_{g}$. We use the luminosity at
5100\AA\ to calculate the bolometric luminosity. From the work of
Eracleous \& Halpern (2003) and Strateva et al. (2003), the inner
radius is about hundreds of $R_{\rm g}$, and the outer radius is
about thousands of $R_{\rm g}$. The outer radius of line-emitting
accretion disk is about near the inner position of torus. We can
assume typical inner and outer radii of $\xi_{1}=450R_{g}$ and
$\xi_{1}=3000R_{g}$ to calculate the energy output for these
double-peaked SDSS AGNs, i.e. $W_{\rm disk}=10^{-1.876}\times
L_{\rm bol} \ \ergs$. The distribution of $\rm {log}(L_{\rm
H\alpha}/W_{\rm disk})$ is $-0.55\pm 0.06$ with the standard
deviation of 0.57 (see also Fig. 8). If assuming as much as $20\%$
of the power was radiated as H$\alpha$ line (dashed line in Fig.
8), our results show that only 36 out of 105 double-peaked AGNs
would generate enough power to produce observed strength of
H$\alpha$ emission. If we adopt the value of 10\%, more objects
(83 out of 105 objects) showed the energy problem. It implied that
the majority of double-peaked AGNs need external illumination of
the disk (e.g. an inner iron torus or corona) to produce the
observed strength of H$\alpha$ line (Strateva et al. 2006; Cao \&
Wang 2006).

\section{conclusions}
We use the simple population synthesis to model the stellar
contributions in double-peaked SDSS AGNs. The reliable stellar
velocity dispersions are obtained for 52 medium-luminous
double-peaked SDSS AGNs with obvious stellar features. We find
that: 1) The black hole mass is from $1.0\times 10^{7} \Msun$ to
$5.5\times 10^{8} \Msun$ and the Eddington ratio is from about
0.01 to about 1;  2) The factor $f$ far deviates from the
virialized value 0.75, suggesting the non-virial dynamics of BLRs;
3) The peak separation is mildly correlated with the Eddington
ratio and SMBH mass with almost the same correlation coefficients,
which can be interpreted in the doppler shift of thin annulus of
BLRs created by gravitational instability; 4) Based on the
line-emitting accretion disk model, we need external illumination
of the accretion disk to produce the observed strength of
H$\alpha$ line. In the future, using different models, we would
fit the double-peaked profiles to constrain the nature of
double-peaked AGNs. We can also use the double-peaked AGNs to
constrain the BLRs origin ( Nicastro 2000; Laor 2003; Bian \& Gu
2007).

\section*{ACKNOWLEDGMENTS}
We are very grateful to the anonymous referee and Ari Laor for
their thoughtful and instructive comments which significantly
improved the content of the paper. We thank Luis C. Ho for  his
very useful comments, and  thank discussions among people in IHEP
AGN group. This work has been supported by the NSFC ( Nos.
10403005, 10473005), the Science-Technology Key Foundation from
Education Department of P. R. China (No. 206053), and and the
China Postdoctoral Science Foundation (No. 20060400502). QSG would
like to acknowledge the financial supports from China Scholarship
Council (CSC) and the NSFC under grants 10221001 and 10633040. JMW
thanks NSFC grants via No. 10325313 and 10521001 and supports from
CAS key project via KJCX2-YW-T03.

Funding for the SDSS and SDSS-II has been provided by the Alfred
P. Sloan Foundation, the Participating Institutions, the National
Science Foundation, the U.S. Department of Energy, the National
Aeronautics and Space Administration, the Japanese Monbukagakusho,
the Max Planck Society, and the Higher Education Funding Council
for England. The SDSS is managed by the Astrophysical Research
Consortium for the Participating Institutions. The Participating
Institutions are the American Museum of Natural History,
Astrophysical Institute Potsdam, University of Basel, Cambridge
University, Case Western Reserve University, University of
Chicago, Drexel University, Fermilab, the Institute for Advanced
Study, the Japan Participation Group, Johns Hopkins University,
the Joint Institute for Nuclear Astrophysics, the Kavli Institute
for Particle Astrophysics and Cosmology, the Korean Scientist
Group, the Chinese Academy of Sciences (LAMOST), Los Alamos
National Laboratory, the Max-Planck-Institute for Astronomy
(MPIA), the Max-Planck-Institute for Astrophysics (MPA), New
Mexico State University, Ohio State University, University of
Pittsburgh, University of Portsmouth, Princeton University, the
United States Naval Observatory, and the University of Washington.

\newpage

\appendix

\section{The relation between the host mass and the SMBH mass}
\begin{figure*}
\begin{center}
\includegraphics[width=9cm]{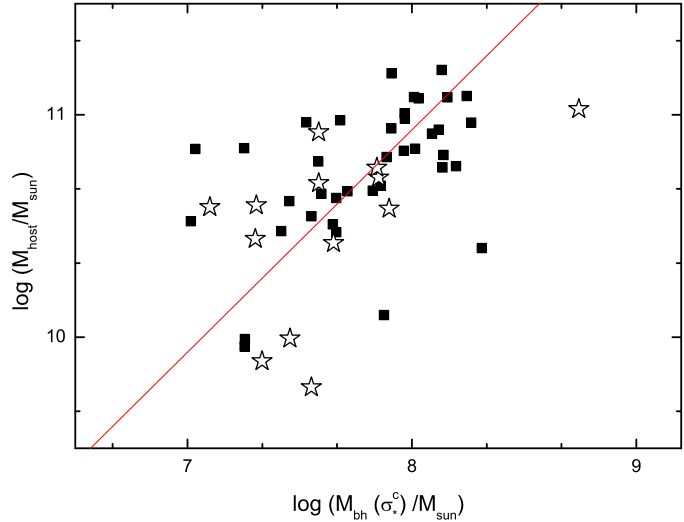}
\caption{The relation between the host mass and the SMBH mass.
With fixed slope of 1, the solid line denotes the best fit with
$R$=0.53 and $p_{\rm null} < 10^{-4}$. Objects with effective S/N
in 4020 \AA\ less then 5 is denoted as open stars.}
\end{center}
\end{figure*}

For the sample of 52 double-peaked SDSS AGNs, assuming $\lambda
L_{\lambda} \propto \lambda^{0.5},\lambda L_{\lambda} (5100\AA)$
is translated to $\lambda L_{\lambda}(5530\AA)$. The host
luminosity in V band is the value of $\lambda
L_{\lambda}(5530\AA)$ multiplied by the stellar fraction at
5530\AA.  We assume that the host luminosity in V band
approximates the bulge luminosity $L_{\rm bulge}$ in V-band. We
then use the following formula to calculate the bulge mass: $\log
(M_{\rm bulge}/\Msun)=1.18\log(L_{\rm bulge} /\Lsun)-1.11$
(Magorrian et al. 1998). Fig. 9 shows this bulge mass versus the
BH mass from $\sigma_*$. We fit with fixed slope as 1, intercept
is $2.93\pm 0.05$, correlation coefficient is 0.53. The null
hypothesis is less then $10^{-4}$. We find that the $M^{\rm
bulge}_V$ versus the $\log \sigma$ relation is consistent with
Fig.4 of Faber et al. 1997. Therefore, the SMBH mass from $L_{\rm
bulge}$ and the Magorrian relation agrees with that from
$\sigma_*$. When we exclude objects with effective S/N in 4020
\AA\ less then 5 (open stars in Fig. 9), for fixed slope of 1, the
best fit gives that intercept is $2.95\pm 0.05$, correlation
coefficient is 0.49.

It is suggested that the mass from the $M_{\rm bh}-\sigma_{*}$
relation could underestimate the SMBH mass for the massive
ellipticals with SMBH mass larger than $10^9\Msun$ (Fig. 2 in
Lauer et al. 2007). For our 52 double-peaked AGNs, their SMBH
masses are all less than $10^9\Msun$ from $M_{\rm bh}-M_{\rm
bulge}$ relation. The present results are less affected by Lauer
et al's findings.


\begin{deluxetable}{lllllllllllll}
\tabletypesize{\tiny} \tablewidth{0pt}
\tablecaption{Results for 52 double-peaked AGNs.}
\startdata
\hline
Name    &  z   &      EW(Ca K) &  $\chi^2$& FC &  S/N  & $\sigma_{*}$ &   $\sigma_{*}^{\rm c}$ & $\sigma_{\rm g}$  &  $\lambda L_{\lambda} (5100\AA)$&  $M_{\rm bh}$ & $L_{\rm bol}/L_{\rm Edd}$  & $\Delta V$\\
      (1) & (2) &(3) & (4) & (5) & (6)&(7)& (8) & (9) & (10)& (11)&(12)&(13)\\
\hline
SDSS J000815.46$-$104620.57  & 0.199 &  $ -3.5 \pm 1.3  $ &  0.86& 0.45 &   4.12 &:148.30 & 146.37& $ 112.2 \pm  4.6  $ & 43.71  &  7.59 & -1.02 & 3100 \\
SDSS J011140.03$-$095834.94  & 0.207 &  $ -3.7 \pm 0.7  $ &  0.91& 0.58 &   3.49 &:110.93 & 110.87& $ 92.5  \pm  31.1 $ & 43.77  &  7.10 & -0.48 & 2300 \\
SDSS J013407.88$-$084129.98  & 0.070 &  $ -7.7 \pm 0.6  $ &  1.07& 0.07 &   9.47 &~121.14 & 121.14$\star$& $ 84.8  \pm  1.3  $ & 42.95  &  7.25 & -1.45 & 3800 \\
SDSS J014901.08$-$080838.23  & 0.210 &  $ -3.1 \pm 7.0  $ &  1.05& 0.64 &   2.78 &:125.25 & 124.48& $ 169.9 \pm  5.3  $ & 43.67  &  7.30 & -0.77 & 2800 \\
SDSS J023253.42$-$082832.10  & 0.265 &  $ -2.7 \pm 0.6  $ &  0.86& 0.30 &   11.00&~186.03 & 182.22& $ 236.1 \pm  14.3 $ & 43.97  &  7.97 & -1.14 & 3200 \\
SDSS J024052.82$-$004110.93  & 0.247 &  $ -3.0 \pm 0.3  $ &  1.31& 0.67 &   7.52 &~121.61 & 121.02& $ 105.9 \pm  3.9  $ & 44.22  &  7.25 & -0.18 & 3100 \\
SDSS J024703.24$-$071421.59  & 0.333 &  $ -4.3 \pm 2.3  $ &  0.81& 0.62 &   3.41 &:293.48 & 284.29& $ 402.1 \pm  19.0 $ & 44.08  &  8.74 & -1.81 & 8100 \\
SDSS J024840.03$-$010032.68  & 0.184 &  $ -4.7 \pm 1.5  $ &  0.72& 0.38 &   6.26 &~155.34 & 153.06& $ 91.9  \pm  0    $ & 43.62  &  7.66 & -1.19 & 2700 \\
SDSS J025220.89$+$004331.32  & 0.170 &  $ -1.3 \pm 0.2  $ &  1.01& 0.50 &   6.81 &~174.65 & 171.41& $ 195.7 \pm  6.9  $ & 43.82  &  7.86 & -1.18 & 2600 \\
SDSS J025951.71$-$001522.78  & 0.102 &  $ -2.5 \pm 0.5  $ &  0.90& 0.62 &   4.51 &:137.38 & 136.00& $ 116.1 \pm  19.3 $ & 43.28  &  7.46 & -1.33 & 3600 \\
SDSS J034931.03$-$062621.05  & 0.287 &  $ -5.1 \pm 0.8  $ &  0.98& 0.61 &   4.32 &:148.31 & 146.38& $ 155.0 \pm  9.2  $ & 44.09  &  7.59 & -0.64 & 5300 \\
SDSS J081700.40$+$343556.34  & 0.062 &  $ -7.6 \pm 1.5  $ &  1.33& 0.16 &   9.35 &~172.92 & 172.92$\star$& $ 200.8 \pm  4.3  $ & 43.08  &  7.88 & -1.94 & 6300 \\
SDSS J081916.28$+$481745.48  & 0.223 &  $ -2.6 \pm 0.4  $ &  1.06& 0.61 &   4.84 &:173.85 & 170.65& $ 170.6 \pm  166.4$ & 43.97  &  7.85 & -1.03 & 5400 \\
SDSS J082133.60$+$470237.33  & 0.128 &  $ -7.9 \pm 2.9  $ &  0.90& 0.19 &   9.47 &~171.11 & 168.05& $ 172.8 \pm  6.5  $ & 43.58  &  7.83 & -1.39 & 5200 \\
SDSS J084535.37$+$001619.52  & 0.260 &  $ -5.2 \pm 1.8  $ &  0.77& 0.33 &   6.52 &~106.66 & 106.82& $ 113.3 \pm  9.5  $ & 43.82  &  7.04 & -0.36 & 5600 \\
SDSS J091459.05$+$012631.30  & 0.198 &  $ -3.2 \pm 0.3  $ &  0.80& 0.33 &   11.38&~156.94 & 154.59& $ 180.7 \pm  15.5 $ & 43.97  &  7.68 & -0.86 & 7400 \\
SDSS J092515.00$+$531711.91  & 0.186 &  $ -3.9 \pm 0.7  $ &  0.79& 0.26 &   8.78 &~203.50 & 198.82& $ 142.7 \pm  11.1 $ & 43.85  &  8.12 & -1.42 & 3400 \\
SDSS J100443.43$+$480156.45  & 0.199 &  $ -5.6 \pm 1.5  $ &  1.00& 0.52 &   3.92 &:172.93 & 169.77& $ 115.2 \pm  3.3  $ & 43.79  &  7.84 & -1.20 & 6600 \\
SDSS J101405.89$+$000620.36  & 0.141 &  $ -7.6 \pm 1.1  $ &  1.50& 0.26 &   12.21&~221.51 & 215.92& $ 166.8 \pm  6.5  $ & 43.85  &  8.26 & -1.56 & 9000 \\
SDSS J103202.41$+$600834.47  & 0.294 &  $ -2.7 \pm 1.0  $ &  0.82& 0.55 &   5.30 &~179.60 & 176.11& $ 164.1 \pm  8.1  $ & 43.98  &  7.91 & -1.08 & 3700 \\
SDSS J104108.18$+$562000.32  & 0.230 &  $ -3.5 \pm 0.9  $ &  1.04& 0.43 &   6.20 &~205.50 & 200.72& $ 181.9 \pm  0.2  $ & 43.82  &  8.14 & -1.47 & 8500 \\
SDSS J104128.60$+$023204.99  & 0.182 &  $ -4.6 \pm 0.5  $ &  0.91& 0.27 &   9.75 &~185.68 & 181.88& $ 191.1 \pm  26.2 $ & 43.78  &  7.96 & -1.33 & 5700 \\
SDSS J104132.78$-$005057.46  & 0.303 &  $ -3.1 \pm 0.3  $ &  1.41& 0.47 &   10.42&~179.70 & 176.20& $ 195.1 \pm  4.2  $ & 44.24  &  7.91 & -0.81 & 4500 \\
SDSS J110742.76$+$042134.18  & 0.327 &  $ -2.5 \pm 0.6  $ &  0.99& 0.42 &   6.12 &~205.09 & 200.32& $ 251.5 \pm  13.6 $ & 44.18  &  8.13 & -1.10 & 3700 \\
SDSS J113021.41$+$005823.04  & 0.132 &  $ -6.9 \pm 1.0  $ &  0.98& 0.31 &   10.85&~148.10 & 146.19& $ 95.2  \pm  0.9  $ & 43.73  &  7.58 & -1.00 & 2100 \\
SDSS J113633.08$+$020747.65  & 0.239 &  $ -4.2 \pm 0.3  $ &  1.18& 0.74 &   6.41 &~143.45 & 141.76& $ 163.2 \pm  10.2 $ & 44.22  &  7.53 & -0.45 & 2500 \\
SDSS J114051.58$+$054631.13  & 0.132 &  $ -11.2\pm 3.9  $ &  0.78& 0.41 &   6.43 &~134.19 & 132.97& $ 138.7 \pm  8.5  $ & 43.50  &  7.42 & -1.06 & 9100 \\
SDSS J115047.48$-$031652.95  & 0.149 &  $ -5.4 \pm 1.1  $ &  0.89& 0.46 &   5.98 &~137.15 & 135.78& $ 91.2  \pm  3.7  $ & 43.69  &  7.45 & -0.91 & 4900 \\
SDSS J122009.55$-$013201.14  & 0.288 &  $ -5.2 \pm 1.2  $ &  0.96& 0.49 &   6.70 &~186.09 & 182.28& $ 179.4 \pm  19.5 $ & 44.02  &  7.97 & -1.09 & 3100 \\
SDSS J130927.67$+$032251.76  & 0.267 &  $ -2.6 \pm 0.5  $ &  1.30& 0.36 &   10.66&~199.92 & 195.42& $ 256.1 \pm  0.5  $ & 43.95  &  8.09 & -1.28 & 3000 \\
SDSS J132442.44$+$052438.86  & 0.116 &  $ -8.8 \pm 10.4 $ &  1.12& 0.45 &   3.20 &:145.40 & 143.62& $ 231.7 \pm  8.9  $ & 42.98  &  7.55 & -1.72 & 9700 \\
SDSS J132834.14$-$012917.64  & 0.151 &  $ -4.2 \pm 1.0  $ &  1.41& 0.60 &   4.70 &:154.22 & 152.00& $ 180.6 \pm  2.2  $ & 43.64  &  7.65 & -1.16 & 3600 \\
SDSS J133312.42$+$013023.73  & 0.217 &  $ -2.8 \pm 0.8  $ &  1.03& 0.55 &   5.18 &~105.37 & 105.59& $ 129.0 \pm  2.3  $ & 43.71  &  7.01 & -0.45 & 3400 \\
SDSS J133338.30$+$041803.94  & 0.202 &  $ -3.0 \pm 0.5  $ &  0.97& 0.36 &   7.34 &~206.10 & 201.29& $ 185.9 \pm  16.7 $ & 43.82  &  8.14 & -1.47 & 4100 \\
SDSS J134617.54$+$622045.47  & 0.116 &  $ -3.4 \pm 0.3  $ &  1.37& 0.64 &   7.94 &~153.91 & 151.71& $ 114.1 \pm  1.3  $ & 43.73  &  7.65 & -1.07 &  --- \\
SDSS J140019.27$+$631426.93  & 0.331 &  $ -2.3 \pm 0.5  $ &  1.05& 0.83 &   5.08 &~204.44 & 199.71& $ 167.1 \pm  25.5 $ & 44.59  &  8.13 & -0.69 & 5300 \\
SDSS J141454.55$+$013358.55  & 0.269 &  $ -4.1 \pm 0.7  $ &  1.15& 0.29 &   10.97&~218.93 & 213.47& $ 287.8 \pm  8.6  $ & 44.03  &  8.24 & -1.36 & 2700 \\
SDSS J141613.37$+$021907.82  & 0.158 &  $ -5.9 \pm 1.2  $ &  1.02& 0.54 &   6.70 &~212.96 & 207.80& $ 117.6 \pm  0.2  $ & 43.81  &  8.20 & -1.54 & 5900 \\
SDSS J141946.06$+$650353.04  & 0.148 &  $ -3.7 \pm 0.3  $ &  1.18& 0.55 &   8.11 &~149.36 & 147.38& $ 154.9 \pm  6.3  $ & 43.85  &  7.60 & -0.90 & 2700 \\
SDSS J142754.76$+$635448.42  & 0.145 &  $ -4.2 \pm 0.3  $ &  1.39& 0.70 &   7.33 &~159.98 & 157.48& $ 117.2 \pm  9.5  $ & 43.86  &  7.71 & -0.99 & 4500  \\
SDSS J143455.31$+$572345.37  & 0.175 &  $ -2.0 \pm 0.4  $ &  1.65& 0.51 &   10.09&~191.25 & 187.18& $ 202.6 \pm  35.9 $ & 44.00  &  8.01 & -1.17 & 3100  \\
SDSS J154534.55$+$573625.12  & 0.268 &  $ -2.2 \pm 0.4  $ &  0.92& 0.46 &   9.34 &~193.04 & 188.88& $ 180.6 \pm  4.9  $ & 44.12  &  8.03 & -1.05 & 2700  \\
SDSS J170102.28$+$340400.60  & 0.094 &  $ -2.7 \pm 0.4  $ &  1.62& 0.75 &   3.79 &:127.55 & 126.66& $ 118.9 \pm  1.4  $ & 43.28  &  7.33 & -1.20 & 1600  \\
SDSS J172102.47$+$534447.29  & 0.192 &  $ -5.0 \pm 2.4  $ &  0.78& 0.35 &   4.45 &:125.66 & 124.86& $ 138.7 \pm  5.6  $ & 43.60  &  7.31 & -0.85 & 3400  \\
SDSS J210109.58$-$054747.31  & 0.179 &  $ -6.2 \pm 1.4  $ &  0.85& 0.32 &   7.52 &~177.42 & 174.04& $ 168.0 \pm  15.1 $ & 43.75  &  7.89 & -1.29 & 3700  \\
SDSS J214935.23$+$113842.04  & 0.239 &  $ -9.5 \pm 5.7  $ &  0.85& 0.56 &   1.92 &:178.69 & 175.24& $ 132.8 \pm  10.0 $ & 43.66  &  7.90 & -1.38 & 10800 \\
SDSS J222132.41$-$010928.76  & 0.288 &  $ -6.3 \pm 1.1  $ &  0.84& 0.56 &   5.01 &~190.63 & 186.59& $ 220.5 \pm  14.9 $ & 44.12  &  8.01 & -1.03 & 8200  \\
SDSS J223302.68$-$084349.13  & 0.058 &  $ -7.2 \pm 0.6  $ &  1.68& 0.37 &   10.05&~121.19 & 121.19$\star$& $ 89.7  \pm  0.9  $ & 43.09  &  7.26 & -1.31 & 2500  \\
SDSS J223336.71$-$074337.10  & 0.174 &  $ -4.6 \pm 0.9  $ &  0.96& 0.36 &   9.12 &~145.42 & 143.64& $ 192.2 \pm  16.4 $ & 43.58  &  7.55 & -1.12 & 3100  \\
SDSS J230545.67$-$003608.55  & 0.269 &  $ -7.1 \pm 2.3  $ &  0.93& 0.35 &   5.52 &~208.17 & 203.25& $ 167.4 \pm  9.8  $ & 44.00  &  8.16 & -1.30 & 8200  \\
SDSS J232721.96$+$152437.31  & 0.046 &  $ -5.2 \pm 0.1  $ &  2.10& 0.45 &   18.59&~221.87 & 221.87$\star$& $ 165.3 \pm  0.4  $ & 43.45  &  8.31 & -2.01 & 4400  \\
SDSS J235128.77$+$155259.15  & 0.096 &  $ -2.8 \pm 0.2  $ &  1.21& 0.66 &   6.96 &~155.30 & 153.03& $ 112.1 \pm  2.9  $ & 43.66  &  7.66 & -1.15 & 3700  \\
\hline
\enddata
\tablecomments{Col. (1): Object name. Col. (2): Redshift. Col.
(3): Equivalent width of Ca K and the error in units of \AA.  Col.
(4): $\chi^2$. Col. (5) fraction of featureless continuum
component. Col. (6): effective S/N from the S/N at 4020 \AA\ and
the fraction of featureless continuum component. Col. (7):
Uncorrected stellar velocity dispersion in units of $\rm km~
s^{-1}$, and values for objects with effective S/N $< 5$ are
preceded by a colon. Col. (8): Corrected stellar velocity
dispersion in units of $\rm km~ s^{-1}$. For 4 objects with Ca II
triplet, it is not need to do the correction, which are shown as
$\star$. Col. (9): Corrected gas velocity dispersion in units of
$\rm km~ s^{-1}$. Col. (10): $\log$ of the monochromatic
luminosity at 5100 \AA\ in units of $\rm ergs~ s^{-1}$ .  Col.
(11): $\log$ of SMBH masses from corrected stellar velocity
dispersion in units of $\rm M_{\odot}$. Col. (12): The Eddington
ratios. Col. (13): The peak separations in units of $\rm km~
s^{-1}$.}
\end{deluxetable}

\end{document}